\documentclass[aps,reprint,amssymb,amsmath,superscriptaddress,notitlepage,prl]{revtex4-1}
\usepackage{graphicx}
\usepackage{miller}
\usepackage[version=4]{mhchem}
\usepackage{xspace}
\usepackage{braket}
\usepackage{siunitx}
\usepackage{hyperref} 
\usepackage{float}
\usepackage{textcomp}
\usepackage{mathrsfs}  
\usepackage{booktabs}
\usepackage{multirow}
\usepackage{silence}
\usepackage{tikz}
\usepackage{bm}
\newcommand{\uvec}[1]{\boldsymbol{\hat{\textbf{#1}}}}
\newcommand*\circled[1]{\tikz[baseline=(char.base)]{
            \node[shape=circle,draw,inner sep=1pt] (char) {#1};}}

\DeclareSIUnit\electrons{e\textsuperscript{--}}
\DeclareSIUnit\neutrons{neutrons}
\DeclareSIUnit\ppm{ppm}
\DeclareSIUnit\ppb{ppb}
\DeclareSIUnit\lines{l}
\DeclareSIUnit{\calorie}{cal}

\newcommand{\oneelec}[1]{\ensuremath{\mathrm{#1}}}

\newcommand{\vect}[1]{\ensuremath{\boldsymbol{#1}}}

\newcommand{\spinstate}[3]{\ensuremath{{}^{#1}\mathrm{#2_{#3}}}}
\newcommand{\vibstate}[2]{\ensuremath{\mathrm{#1_{#2}}}}

\newcommand{\NVnb}{\ensuremath{\mathrm{NV}}\xspace}

\newcommand{\NVminus}{\ensuremath{\NVnb^{-}}\xspace}

\newcommand{\SiVneutral}{\ensuremath{\mathrm{SiV}^{0}}\xspace}
\newcommand{\SiVminus}{\ensuremath{\mathrm{SiV}^{-}}\xspace}
\newcommand{\SiV}{\ensuremath{\mathrm{SiV}}\xspace}

\newcommand{\GeVminus}{\ensuremath{\mathrm{GeV}^{-}}\xspace}

\newcommand{\boltzm}{\ensuremath{k_{B}}\xspace}

\newcommand{\Tetrahedral}{\ensuremath{\mathrm{T}_\mathrm{d}}\xspace}
\newcommand{\Trigonal}{\ensuremath{\mathrm{C}_{\mathrm{3v}}}\xspace}
\newcommand{\DIIId}{\ensuremath{\mathrm{D_{3d}}}\xspace}
\newcommand{\Triclinic}{\ensuremath{\mathrm{C_s}}\xspace}

\newcommand{\stress}{\ensuremath{\sigma}}

\setcitestyle{numbers,square}

\AtBeginDocument{
\heavyrulewidth=.08em
\lightrulewidth=.05em
\cmidrulewidth=.03em
\belowrulesep=.65ex
\belowbottomsep=0pt
\aboverulesep=.4ex
\abovetopsep=0pt
\cmidrulesep=\doublerulesep
\cmidrulekern=.5em
\defaultaddspace=.5em
}

\begin{document}

\title{Electronic structure of the neutral silicon-vacancy center in diamond}

\author{B.\ L.\ Green}
\email{b.green@warwick.ac.uk}
\altaffiliation{Corresponding Author}
\affiliation{Department of Physics, University of Warwick, Coventry, CV4 7AL, UK}
\affiliation{EPSRC Centre for Doctoral Training in Diamond Science and Technology, UK}
\author{M.\ W.\ Doherty}
\affiliation{Laser Physics Centre, Research School of Physics and Engineering, Australian National University, Australian Capital Territory 2601, Australia}
\author{E.\ Nako}
\affiliation{Department of Physics, University of Warwick, Coventry, CV4 7AL, UK}
\affiliation{EPSRC Centre for Doctoral Training in Diamond Science and Technology, UK}
\author{N.\ B.\ Manson}
\affiliation{Laser Physics Centre, Research School of Physics and Engineering, Australian National University, Australian Capital Territory 2601, Australia}
\author{U.\ F.\ S.\ D'Haenens-Johansson}
\affiliation{Gemological Institute of America, 50 W 47\textsuperscript{th} St, New York, NY 10036, USA}
\author{S.\ D.\ Williams}
\author{D.\ J.\ Twitchen}
\affiliation{Element Six Limited, Global Innovation Centre, Fermi Avenue, OX11 0QR, UK}
\author{M.\ E.\ Newton}
\affiliation{Department of Physics, University of Warwick, Coventry, CV4 7AL, UK}
\affiliation{EPSRC Centre for Doctoral Training in Diamond Science and Technology, UK}

\begin{abstract}
The neutrally-charged silicon vacancy in diamond is a promising system for quantum technologies that combines high-efficiency, broadband optical spin polarization with long spin lifetimes ($T_2\approx\SI{1}{\milli\second}$ at \SI{4}{\kelvin}) and up to \SI{90}{\percent} of optical emission into its \SI{946}{\nano\meter} zero-phonon line. However, the electronic structure of \SiVneutral{} is poorly understood, making further exploitation difficult. Performing photoluminescence spectroscopy of \SiVneutral{} under uniaxial stress, we find the previous excited electronic structure of a single \spinstate{3}{A}{1u} state is incorrect, and identify instead a coupled $\spinstate{3}{E}{u}-\spinstate{3}{A}{2u}$ system, the lower state of which has forbidden optical emission at zero stress and so efficiently decreases the total emission of the defect: we propose a solution employing finite strain to form the basis of a spin-photon interface. Isotopic enrichment definitively assigns the \SI{976}{\nano\meter} transition associated with the defect to a local mode of the silicon atom. 
\end{abstract}

\maketitle

Optically-accessible solid state defects are promising candidates for scalable quantum information processing \cite{Aharonovich2016,Rogers2014b}. Diamond is the host crystal for two of the most-studied point defects: the negatively-charged nitrogen vacancy (\NVminus{}) center \cite{Doherty2013}, and the negatively-charged silicon vacancy (\SiVminus{}) center \cite{Rogers2014a}. \NVminus{} has been successful in a broad range of fundamental \cite{Gaebel2006,Hensen2015} and applied \cite{Maletinsky2012,Wang2015b,Rondin2014} quantum experiments, with spin-photon \cite{Togan2010} and spin-spin \cite{Bernien2013} entanglement protocols well-established. The superior photonic performance of \SiVminus{}, with \SI{>70}{\percent} of photonic emission into its zero phonon line (ZPL), has enabled it to make a rapid impact in photonic quantum platforms \cite{Sipahigil2016,Sipahigil2014}. However, \SiVminus{} possesses poor spin coherence lifetimes due to phononic interactions in the ground state (GS) \cite{Meesala2018}, requiring temperatures of \SI{<100}{\milli\kelvin} to achieve $T_2\approx\SI{400}{\micro\second}$ without decoupling \cite{Sukachev2017}. 

\begin{figure}
	\centering
	\includegraphics[width=\columnwidth]{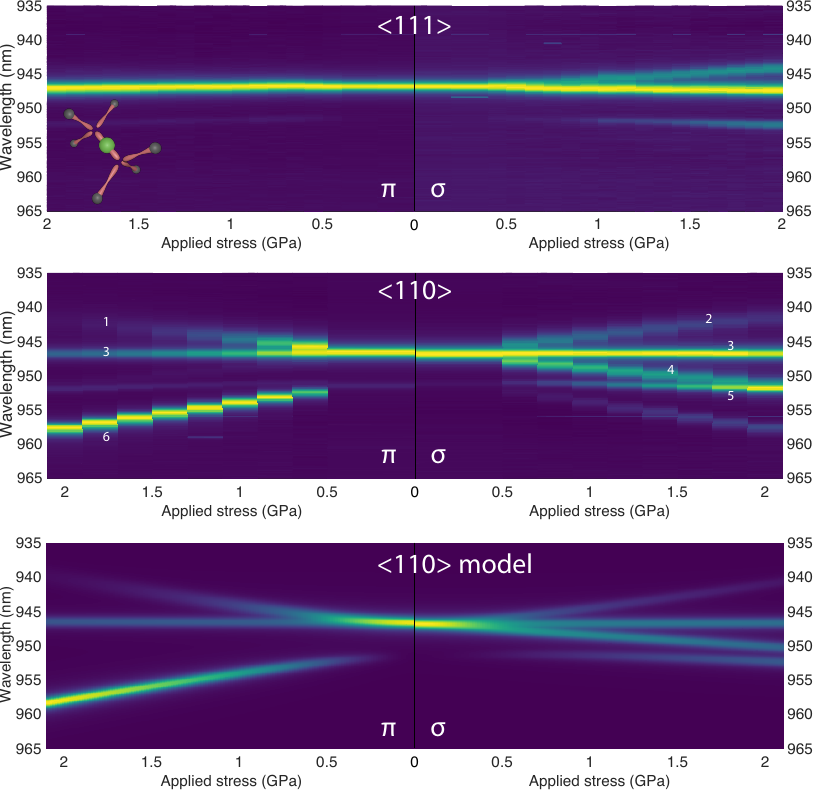}
	\caption{\SiVneutral{} photoluminescence spectra at \SI{80}{\kelvin} as a function of applied stress along \hkl<111> (top) and \hkl<110> (middle). In each case, $\pi$ ($\sigma$) indicates detection polarization parallel (perpendicular) to the stress direction. The transition at \SI{946}{\nano\meter} splits into components 1--4 under \hkl<110> stress, with thermalisation between the components observed at high stress indicating electronic degeneracy. A pair of stress-induced transitions (5,6) originate at approximately \SI{951}{\nano\meter}. Inset, top: the geometric form of \SiVneutral{}, with the Si atom on-axis in the split-vacancy configuration. Bottom: simulation of the \hkl<110> stress spectra using the model described in main text.}
	\label{fig:stress_surfaces}
\end{figure}

Recent work on \SiVneutral{}, the neutrally-charged silicon vacancy in diamond, has demonstrated that it combines high-efficiency optical spin polarization \cite{Green2017c} with long spin lifetimes ($T_2\approx\SI{1}{\milli\second}$ at \SI{4}{\kelvin} \cite{Rose2017}) and a high degree of coherent emission: the defect potentially possesses the ideal combination of \SiVminus{} and \NVminus{} properties. Exploitation of these promising properties is hindered by poor understanding of the defect's electronic structure. Electron paramagnetic measurements (EPR) of \SiVneutral{} indicate it has a spin triplet \spinstate{3}{A}{2g} GS and \DIIId{} symmetry \cite{Edmonds2008a}, with the silicon atom residing on-axis in a split-vacancy configuration [Fig~\ref{fig:stress_surfaces}, inset]. Optically-excited EPR measurements directly relate the \SiVneutral{} spin system to a zero phonon line (ZPL) at \SI{946}{\nano\meter} \cite{Green2017c}: optical absorption experiments and density functional theory (DFT) calculations have assigned the ZPL excited state (ES) to \spinstate{3}{A}{1u} symmetry \cite{DHaenens-Johansson2011,Gali2013}. Temperature-dependent PL measurements indicate the presence of an optically-inactive state below the luminescent excited state \cite{DHaenens-Johansson2011}.

The advances in exploitation of \NVminus{} and \SiVminus{} have been driven by a concerted effort in the fundamental understanding of the physics of the centers themselves. In this Letter, we employ photoluminescence (PL) spectroscopy to study an ensemble of \SiVneutral{} under applied uniaxial stress, and show that the previous assignment of a single excited state \spinstate{3}{A}{1u} is incorrect. We find that the \SI{946}{\nano\meter} excited state is \spinstate{3}{E}{u}, with a \spinstate{3}{A}{2u} state approximately \SI{6.8}{\milli\electronvolt} below it. The latter transition is forbidden by symmetry at zero stress and therefore efficiently reduces the emission intensity of unstrained \SiVneutral{} centers at low temperature. However, under finite strain, the proposed electronic structure enables the possibility of resonantly exciting spin-selective optical transitions between the \spinstate{3}{A}{2g} GS and \spinstate{3}{A}{2u} ES. The latter state is shown definitively to participate in the optical spin polarization mechanism of \SiVneutral{}. Finally, we demonstrate that the \SI{976}{\nano\meter} transition associated with \SiVneutral{} \cite{Green2017c}, previously hypothesised to be a strain-induced transition \cite{Gali2013}, is actually a pseudo-local vibrational model (LVM) of \SiVneutral{} primarily involving the silicon atom. 

We apply uniaxial stress to a diamond crystal grown by chemical vapour deposition: the crystal was doped with silicon during growth to create \SiVminus{} and \SiVneutral{}. Uniaxial stress was applied to the sample using a home-built ram driven by pressurized nitrogen gas. PL measurements were collected under excitation at \SI{785}{\nano\meter} as a function of applied stress in both the \hkl<111> and \hkl<110> directions (see \footnote{\label{footnote:suppInfo}See Supplemental Material at http://abc for description of the experimental geometry and apparatus, comparison of spectra with different input polarizations, derivation of the analytical solutions to the coupled stress Hamiltonian, the model parameters used to generate the simulation, and detail on the computation of transition intensities.} for detail). We measured spectra for all four combinations of excitation and detection polarization parallel ($\pi$) and perpendicular ($\sigma$) to the stress axis. We found that the spectra are essentially invariant to excitation polarization \cite{Note1}. This is likely due to the excitation mechanism being polarization-insensitive photoionization, as our \SI{785}{\nano\meter} (\SI{1.58}{\electronvolt}) excitation laser is above the \SI{830}{\nano\meter} (\SI{1.50}{\electronvolt}) photoionization threshold of \SiVneutral{} \cite{Allers1995}. We can thus focus on analysing just the spectra for the two detection polarizations ($\pi$, $\sigma$) arising from a single excitation polarization ($\pi$).

The problem of uniaxial stress applied to a trigonal defect in a cubic crystal has been described several times \cite{Hughes1967,Davies1976a,Rogers2015}, so we summarise the results for transitions to an orbital singlet GS, as found in \SiVneutral{}. In both \hkl<111> and \hkl<110> applied stress, the orientational degeneracy of the defect is lifted into two classes of orientation, classified by the angle between their high-symmetry axis and the uniaxial stress axis. For an orbital singlet-to-singlet ($A\leftrightarrow A$) transition, only one transition per orientation is possible: when taking into account both orientation classes, we expect a maximum of two transitions per spectrum. In the orbital singlet-to-doublet ($A\leftrightarrow E$) case, two transitions per orientation are possible, leading to a maximum of four transitions per spectrum. \hkl<111> stress does not remove the electronic degeneracy of the $E_{x},E_{y}$ orbitals for the orientation parallel to the applied stress, and hence a maximum of three transitions are expected. 

For uniaxial stress applied along the \hkl<111> axis, the \SI{946}{\nano\meter} ZPL splits into three transitions, two of which are almost degenerate but which possess different emission polarization [Fig.~\ref{fig:stress_surfaces}]. This is consistent with the $A\leftrightarrow E$ case described earlier. Under \hkl<110> uniaxial stress, we identify four distinct components originating at the ZPL, again consistent with an $A\leftrightarrow E$ transition. The intensities of the different components varies as a function of applied stress, confirming the presence of electronic degeneracy in the excited state. For both stress directions, we observe additional lower-energy transitions originating at $\approx\SI{951}{\nano\meter}$: the transitions gain intensity as a function of stress [Fig.~\ref{fig:stress_surfaces}]. We measure only two components, indicating the presence of an additional orbital singlet state. At a constant applied stress of $\sigma_{\hkl<110>}=\SI{1.3}{\giga\pascal}$, decreasing the temperature increases the intensity of the stress-induced transitions at the expense of the ZPL transitions. Therefore, we conclude the additional $A$ state lies below the excited $E$ state, rather than above the ground \spinstate{3}{A}{2g}.

\begin{figure*}[t]
	\centering
	\includegraphics[width=0.8\textwidth]{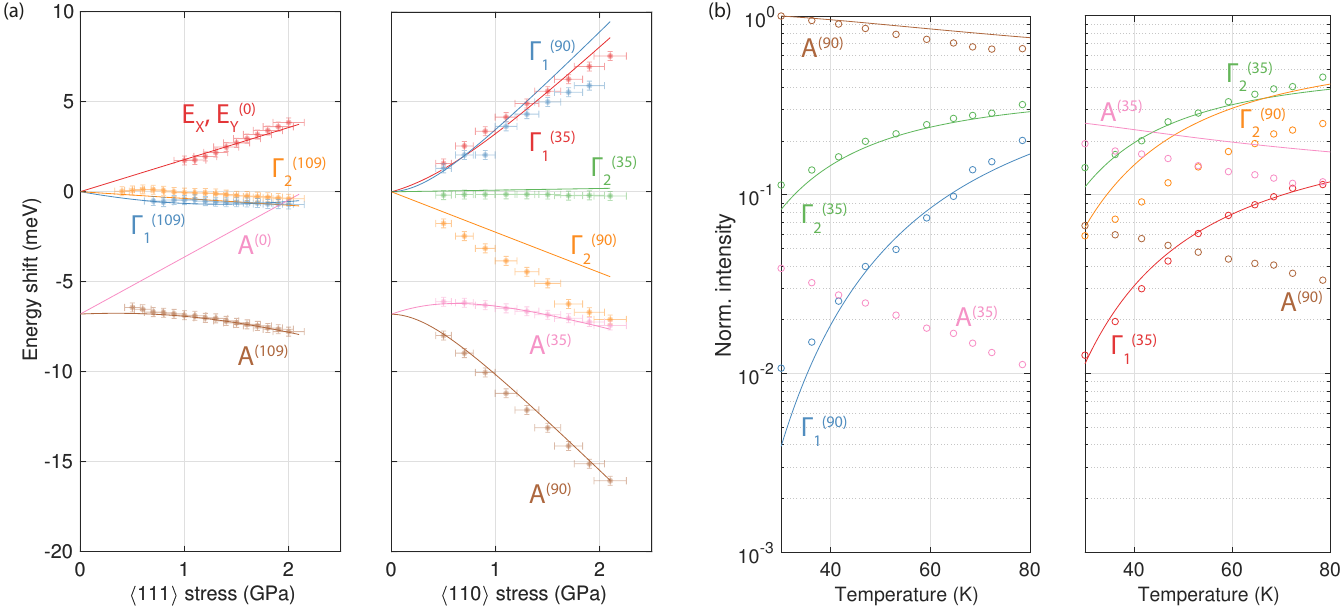}
	\caption{Comparison of experimental data (dots) with the coupled $E-A$ model (solid lines). Transitions are labelled with the state ($A$ and $\Gamma_1$, $\Gamma_2$ for the $E$ state) and the angle between the symmetry axis of the sub-ensemble and the stress axis (in degrees). (a) Transition energies as a function of applied stress in the \hkl<111> (left) and \hkl<110> (right) directions. Theoretical intensity of the $A^{(0)}$ transition is 0, and the line was not observed in experiment. (b) Transition intensities at an applied \hkl<110> stress of \SI{1.3}{\giga\pascal} as a function of sample temperature. Data are given in both $\pi$ (left) and $\sigma$ (right) detection polarizations. The data have been normalized to the most intense transition.}
	\label{fig:stress_and_intensity}
\end{figure*}

In order to construct a model of the excited state behavior, we must establish the origin of the lower-energy $A$ state. There are three possible origins: (1) spin-orbit (SO) fine structure arising from the $E$ level; (2) Jahn-Teller (JT) vibronic structure arising from the $E$ level; and (3) a totally independent $A$ level. An SO interaction of \SI{6.5}{\milli\electronvolt} ($\approx\SI{1.57}{\tera\hertz}$) is inconsistent with the magnitude of the SO interaction in \SiVminus{} ($\SI{250}{\giga\hertz}$ \cite{Rogers2014a}) and \GeVminus{} (\SI{1.06}{\tera\hertz} \cite{Palyanov2015}) and would yield additional $A$ and $E$ states (as in \NVminus{} ES \cite{Delaney2010}) and hence we reject this possibility. A JT distortion would place the $A$ state above the $E$ and hence is inconsistent with experiment. Additionally, the piezospectroscopic parameters describing the singlet and doublet states are significantly different \cite{Note1}, as would be expected if they arise from distinct electronic states \cite{Davies1979}. We conclude that the singlet is an additional electronic state and is not derived from the doublet. Experimentally, we find the singlet transitions are polarized in pure $\sigma$ for \hkl<111> stress, and pure $\sigma$, $\pi$ for \hkl<110> stress [Fig.~\ref{fig:stress_surfaces}]: this identifies the $A$ level as possessing $\Gamma_1$ symmetry in the lowered \Triclinic symmetry of the defect under stress \cite{Note1}. 

Building on previous numerical descriptions of a coupled $E-A$ system in trigonal symmetry \cite{Davies1979}, we construct a full analytical treatment of this problem. For a given SiV sub-ensemble under applied stress, the coupled Hamiltonian is 
\begin{equation}
H=\left(
\begin{array}{ccc}
 	W+\alpha' 	& \gamma ^c 		& \beta ^c \\
 	\gamma ^c 	& \alpha +\beta  	& \gamma  \\
 	\beta ^c 	& \gamma  			& \alpha -\beta  \\
\end{array}
\right)
\end{equation}%
where $\alpha$, $\beta$, $\gamma$ ($\alpha^\prime$) describe the response to stress of the $E$ ($A$) state, $\beta^c$ and $\gamma^c$ describe coupling between the two states, and $W$ is the energy difference between the states at zero stress. $\alpha^{(\prime)}$, $\beta^{(c)}$ and $\gamma^{(c)}$ are functions of the state-dependent piezospectroscopic parameters and are linear in applied stress. The eigenenergies of this Hamiltonian can be parameterised as follows (see \cite{Note1} for derivation)
\begin{align}
\begin{split}
E(A)= & \frac{1}{2}\left(\alpha +\Delta +W+\alpha '\right)\nonumber\\
	& -\frac{1}{2}\left[\left(\alpha +\Delta -W-\alpha '\right)^2+4\Omega ^2\right]^{1/2}\nonumber
\end{split}\\
\begin{split}\label{eqn:hamiltonian_solutions}
E(\Gamma _1) = & \frac{1}{2}\left(\alpha +\Delta +W+\alpha '\right)\\
& +\frac{1}{2}\left[\left(\alpha +\Delta -W-\alpha '\right)^2+4\Omega ^2\right]^{1/2}
\end{split}\\
E(\Gamma _2) = &\alpha -\Delta\nonumber
\end{align}%
where $\Delta$ is the stress splitting of the $E$ level in the absence of the coupling to the $A$ level and $\Omega$ is the coupling between the $A$ level and the $E$ state that also has $\Gamma _1$ symmetry under $C_s$ stress. The intensities of the corresponding lines in detection polarization $p$ are
\begin{align}
I_p(A)&=Z^{-1}e^{-E(A)\left/k_B\right.T}I_{1p}\sin ^2\frac{\phi }{2}\nonumber\\
I_p\left(\Gamma _1\right)&=Z^{-1}e^{-E\left(\Gamma _1\right)/k_BT}I_{1p}\cos ^2\frac{\phi }{2}\\
I_p\left(\Gamma _2\right)&=Z^{-1}e^{-E\left(\Gamma _2\right)/k_BT}I_{2p}\nonumber
\end{align}%
where $I_{1p}$ and $I_{2p}$ are intensities of $p$-polarization components of the $\Gamma _1$ and $\Gamma _2$ transitions (given in \cite{Note1}), $\phi =\arctan \frac{2\Omega }{\alpha +\Delta -W-\alpha '}$ is the angle describing the coupling between the $A$ and the $\Gamma _1$ substate of the $E$ state, and $Z$ is the partition function.

The result of a least-squares fit of this model simultaneously to the experimental \hkl<110> and \hkl<111> spectra as a function of stress is given in Fig.~\ref{fig:stress_and_intensity}(a): piezospectroscopic parameters are detailed in the SI \cite{Note1}). The output of the model was tested by comparing it to the transition intensities of spectra measured as a function of temperature at a fixed $\sigma_{\hkl<110>}=\SI{1.3}{\giga\pascal}$ [Fig.~\ref{fig:stress_and_intensity}(b)]. The ordering and behavior of all transitions matches the experiment and hence we accept the coupled $E-A$ model as a suitable description of the \SiVneutral{} excited state. 

There are several reasons why the model fit is not perfect. Intrinsic inhomogeneous stress will introduce non-linearities into the line-shifts at low stress; small misalignments or non-uniaxial stress will modify the shift-rates from those taken into account by the model, which will be exacerbated if these effects are different in the two stress directions. Finally, Jahn-Teller interactions in the $E$ state, and pseudo-Jahn Teller interactions between the $E$ and $A$ are not taken into account within the model: high quality absorption data under stress are required to confirm the presence of these interactions, and the low concentration of \SiVneutral{} in the present sample prohibits absorption measurements.

With the excited states' orbital degeneracy and symmetry under stress confirmed, we now reconcile our observations with the electronic model of \SiVneutral{}. The EPR-active \spinstate{3}{A}{2g} GS arises from the molecular orbital (MO) configuration \oneelec{a_{1g}^2{}}\oneelec{a_{2u}^{2}}\oneelec{e_u^{4}}\oneelec{e_g^2} ($\equiv\oneelec{e_g^2}$ in the hole picture, used henceforth), along with \spinstate{1}{E}{g}, \spinstate{1}{A}{1g} \cite{Gali2013}. The previously-assigned \spinstate{3}{A}{1u} ES arises from \oneelec{e_u^1e_g^1} \cite{DHaenens-Johansson2011}, in addition to \spinstate{1}{A}{1u}, \spinstate{1}{A}{2u}, \spinstate{1}{E}{u}, \spinstate{3}{A}{2u} and \spinstate{3}{E}{u} states. As \oneelec{e_g^2} and \oneelec{e_u^1e_g^1} are the two lowest-energy one-electron configurations \cite{Gali2013}, we identify the doubly-degenerate ES observed under stress with the \spinstate{3}{E}{u} (\oneelec{e_u^1e_g^1}) state. 

The requirement of applied stress for observation of the singlet transitions [Fig.~\ref{fig:stress_surfaces}] indicates that the transitions are forbidden by orbital symmetry but not spin. As the only $S=1$ state arising from the \oneelec{e_g^2} configuration, we assume that the GS of this transition is the EPR-active \spinstate{3}{A}{2g}: the singlet is then restricted by symmetry selection rules to \spinstate{3}{A}{1g}, \spinstate{3}{A}{2g} and \spinstate{3}{A}{2u}. The observed $\Gamma_1$ symmetry under stress may be derived from both \spinstate{}{A}{1g} and \spinstate{}{A}{2u} in \DIIId{}; however, only the latter is consistent with the electronic model and hence we assign the symmetry \spinstate{3}{A}{2u} (\oneelec{e_u^1e_g^1}). We identify this state with the \SI{\approx 5}{\milli\electronvolt} state observed in temperature-dependent PL measurements, where the intensity of the ZPL was shown to decrease with decreasing temperature \cite{DHaenens-Johansson2011}. 

\begin{figure}
	\centering
	\includegraphics[width=0.9\columnwidth]{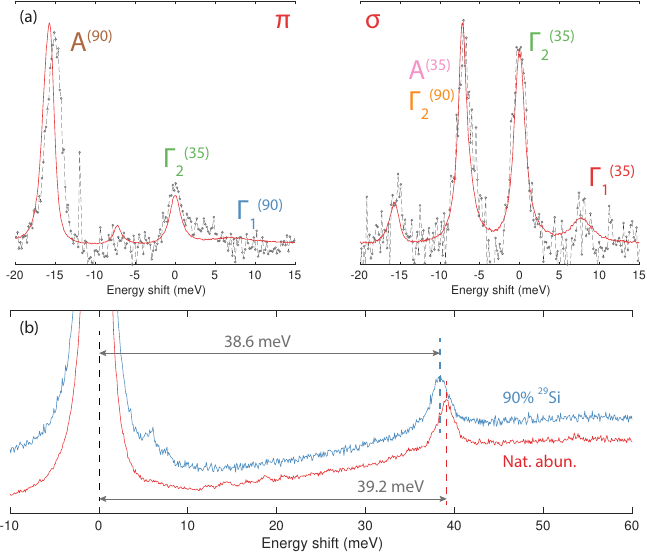}
	\caption{(a) Comparison of PL measurements of the \SI{946}{\nano\meter} and associated stress-induced transitions (solid lines) with the \SI{976}{\nano\meter} local mode (dots). Measurements collected at $\stress_{\hkl<110>}=\SI{2.1}{\giga\pascal}$ for both $\pi$ (left) and $\sigma$ detection polarization. Individual transitions are labelled as in Fig.~\ref{fig:stress_and_intensity}. (b) Effect of isotopic enrichment on the \SI{976}{\nano\meter} local vibrational mode. The mode shifts from $\Omega_0 = \SI{39.2}{\milli\electronvolt}$ in natural abundance material (\SI{92}{\percent} \ce{^{28}Si}) to $\Omega^{\ast}=\SI{38.6}{\milli\electronvolt}$ in a sample enriched with \SI{90}{\percent} \ce{^{29}Si}. Treating the mode as a simple harmonic oscillation of the silicon atom yields $\Omega^{\ast} = \SI{38.6}{\milli\electronvolt}$, matching experiment. ZPLs have been fixed at zero for clarity.}
	\label{fig:976_figure}
\end{figure}

In addition to the purely electronic transitions discussed above, the PL spectrum of \SiVneutral{} also exhibits a small feature at \SI{976}{\nano\meter} \cite{Green2017c}. In our measurements, we find that the energy shift of the transition under stress is essentially identical \num{946} and \SI{951}{\nano\meter} transitions [Fig.~\ref{fig:976_figure}(a)] \cite{Note1}. As the line is at lower energy than the associated ZPLs we associate it with a pseudo-LVM in the common GS. This observation is incompatible with previous density functional theory (DFT) calculations suggesting that this transition is a stress-induced electronic transition between a \spinstate{3}{E}{g} ES and the \spinstate{3}{A}{2g} GS \cite{Gali2013}. 

To investigate the participation of Si in the pseudo-LVM, PL measurements of a sample grown with isotopically enriched silicon dopant were performed: we find that the vibration frequency drops from \SI{39.2}{\milli\electronvolt} in a natural abundance sample ($\SI{>90}{\percent}$ \ce{^{28}Si}) to \SI{38.6}{\milli\electronvolt} in a sample enriched with $\SI{90}{\percent}$ \ce{^{29}Si} [Fig.~\ref{fig:976_figure}(b)]. Modelling the vibration as a simple harmonic oscillator, the mode frequency under isotopic enrichment is given by $\Omega^{\ast} = \Omega_0 \sqrt{m^{\ast}/m_{\text{0}}}$, where $m^{\ast}$ is the effective mass of the isotopic enrichment, and $\Omega_0$, $m_0$ are the mode frequency and effective mass in a natural abundance sample, respectively. Applying this model yields $\Omega^{\ast}_{\mathrm{model}}=\SI{38.6}{\milli\electronvolt}$, matching the experimental value. This confirms that the LVM is primarily due to oscillation of the Si within the vacancy `cage', and is only weakly coupled to the bulk. Finally, the symmetry of the LVM may be addressed. The similar polarization behavior of the \num{946} and \SI{976}{\nano\meter} transitions [Fig~\ref{fig:976_figure}(a)] indicates an \vibstate{a}{1g} mode. However, only \vibstate{e}{u} or \vibstate{a}{2u} silicon oscillations participate in pseudo-LVM modes \cite{Londero2018}: in both these cases, the overall mode symmetry $\spinstate{3}{A}{2g}\otimes\Gamma_{\text{LVM}}$ becomes \textit{ungerade} and thus vibronic transitions from both \spinstate{3}{E}{u} and \spinstate{3}{A}{2u} excited states are forbidden by parity. We may reconcile the spectroscopic data with the model only by considering symmetry-lowering distortions. For example, under instantaneous symmetry-lowering distortions from $\DIIId{}\rightarrow\Trigonal$ due to (pseudo-)Jahn-Teller distortions in the ES, the \vibstate{a}{2u} mode becomes \vibstate{a}{1} and the vibronic transition is no longer forbidden. We observe no sharp mode related to the \vibstate{e}{u} oscillation of the silicon. A similarly complex situation is encountered in \SiVminus{}, where two pseudo-LVMs have been identified at \num{40} and \SI{64}{\milli\electronvolt} \cite{Rogers2014a}. Studies of the latter indicate that its frequency is well-approximated by a simple harmonic oscillator model \cite{Dietrich2014} and essentially involves only the silicon atom, as we find for the \SI{39}{\milli\electronvolt} mode of \SiVneutral{}. However, experimental measurements assign the \SI{64}{\milli\electronvolt} mode to \vibstate{a}{2u} symmetry \cite{Rogers2014,Dietrich2014} through polarized single-center studies, whereas recent hybrid-DFT calculations assign the mode \vibstate{e}{u} symmetry and argue that the \SI{40}{\milli\electronvolt} mode is not an LVM \cite{Londero2018}. Further work is required to definitively identify the vibrational states of \SiV{} in both charge states. 

\begin{figure}
	\centering
	\includegraphics[width=\columnwidth]{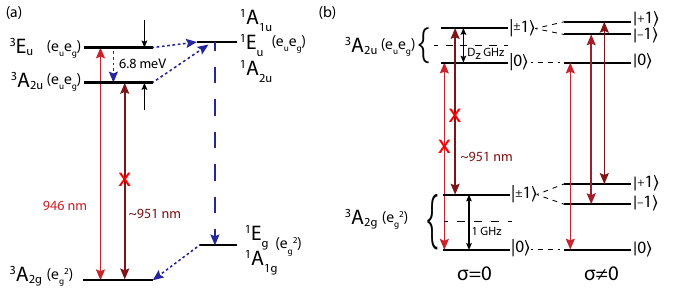}
	\caption{(a) The electronic structure of \SiVneutral{} proposed as a result of uniaxial stress measurements. The ordering and relative energies of the spin singlets is not known. Electronic configurations are described in the hole picture i.e. $\oneelec{e_ue_g}\equiv\oneelec{a_{2u}^2a_{1g}^2e_u^3e_g^3}$. (b) Proposed scheme for spin-dependent initialization and readout of the \SI{951}{\nano\meter} transition under a small applied strain. $D_z$ is not known.}
	\label{fig:electronic_states}
\end{figure}

With knowledge of the excited state symmetries and behavior under stress, we may re-analyse recent measurements of the spin polarization behavior \cite{Green2017c,Rose2017}. The latter measurement identifies significant spin polarization at approximately \SI{951}{\nano\meter} (Fig.~S9 \cite{Rose2017}): in light of our new results on the stress-induced optical transition at \SI{951}{\nano\meter}, we understand that the measurement was performed on a strained ensemble, and interpret its visibility in an absorption spectrum as a direct transition from the \spinstate{3}{A}{2g} ground state to the \spinstate{3}{A}{2u} state [Fig.~\ref{fig:electronic_states}(a)]. As the measurements were completed by reading out spin polarization from the \spinstate{3}{A}{2g} GS, this is direct evidence that the \spinstate{3}{A}{2u} ES is involved in the spin polarization mechanism. At \SI{4}{\kelvin}, $\boltzm{}T\approx\SI{0.3}{\milli\electronvolt}$ and hence thermal excitation from \spinstate{3}{A}{2u} to \spinstate{3}{E}{u} is negligible. The spin polarization mechanism must therefore either involve interactions with both the \spinstate{3}{E}{u} and \spinstate{3}{A}{2u} states, or via phonon relaxation from the \spinstate{3}{E}{u} state through \spinstate{3}{A}{2u} [Fig~\ref{fig:electronic_states}(a)]. Information on the relative ordering of the singlet states is required for a full description of the spin polarization mechanism \cite{Note1}.

The thermal interaction of the \spinstate{3}{E}{u} and \spinstate{3}{A}{2u} states poses a problem for the use of \SiVneutral{} as a photonic resource, as the intensity of the \SI{946}{\nano\meter} transition decreases with decreasing temperature due to thermal depopulation from \spinstate{3}{E}{u} into \spinstate{3}{A}{2u}: typically, \SI{<20}{\kelvin} is required to isolate spin-conserving optical transitions in diamond \cite{Fu2009,Nicolas2018}. For small ($\lesssim\SI{0.3}{\giga\pascal}$) stresses applied perpendicular to the symmetry axis, the intensity and frequency of the \SI{951}{\nano\meter} transition is quadratic in stress: the stress will also remove the $m_s=\pm1$ spin degeneracy in the spin triplets. Under stress, the spin-conserving optical transitions between \spinstate{3}{A}{2g} GS and \spinstate{3}{A}{2u} ES are no longer forbidden [Fig.~\ref{fig:electronic_states}(b)], and in conjunction with the spin polarization mechanism in \SiVneutral{} may enable spin-dependent optical initialization and readout at low magnetic field. To resolve spin-dependent optical transitions, we require the difference in the zero-field splitting of the GS and ES to be larger than the inhomogeneous linewidth of the transitions themselves. Implementation of this scheme would form the foundation of an \SiVneutral{} spin-photon interface \cite{Togan2010}. Future work should include monitoring strained \SiVneutral{} centers in both EPR and resonant PL to determine the effect of strain on the spin-spin interactions in both the orbital singlet states, and measurement of single centers under strain to identify spin-conserving optical transitions.

\begin{acknowledgments}
We thank B. G. Breeze at the University of Warwick Spectroscopy Research Technology Platform for helpful discussion and assistance with experiments. BLG gratefully acknowledges the financial support of the Royal Academy of Engineering. This work is supported by EPSRC Grants No.\ EP/L015315/1 and EP/M013243/1, and ARC Grants No.\ DE170100169 and DP140103862.
\end{acknowledgments}

\clearpage
\widetext
\begin{center}
\textbf{\large Supplemental Material}
\end{center}
\setcounter{equation}{0}
\setcounter{figure}{0}
\setcounter{table}{0}
\setcounter{page}{1}
\makeatletter
\renewcommand{\theequation}{S\arabic{equation}}
\renewcommand{\thefigure}{S\arabic{figure}}
\renewcommand{\thetable}{S\arabic{table}}
\renewcommand{\thepage}{S\arabic{page}}
\renewcommand{\bibnumfmt}[1]{[S#1]}
\renewcommand{\citenumfont}[1]{S#1}

\section{Experimental detail}
\begin{figure}[hbt]
	\centering
	\includegraphics[width=0.3\textwidth]{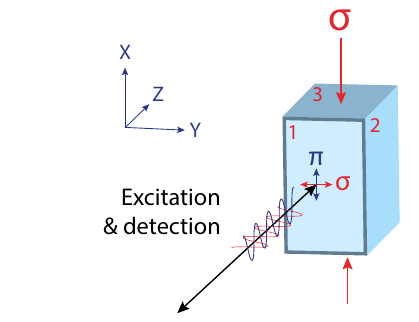}
    \caption{Geometry for stress experiments: the excitation / detection are backscattered for all measurements. Faces 1, 2, 3 are $\hkl[1-1-1]$, $\hkl[-11-2]$, $\hkl[110]$ ($\hkl[-11-2]$, $\hkl[110]$, $\hkl[1-1-1]$) for $\hkl<110>$ ($\hkl<111>$) stress, respectively. The electric field vector for excitation and detection is either parallel ($\pi$) or perpendicular ($\sigma$) to the stress axis.}
    \label{fig:stress_schematic}
\end{figure}

\begin{figure}[h]
	\centering
	\includegraphics[width=0.8\textwidth]{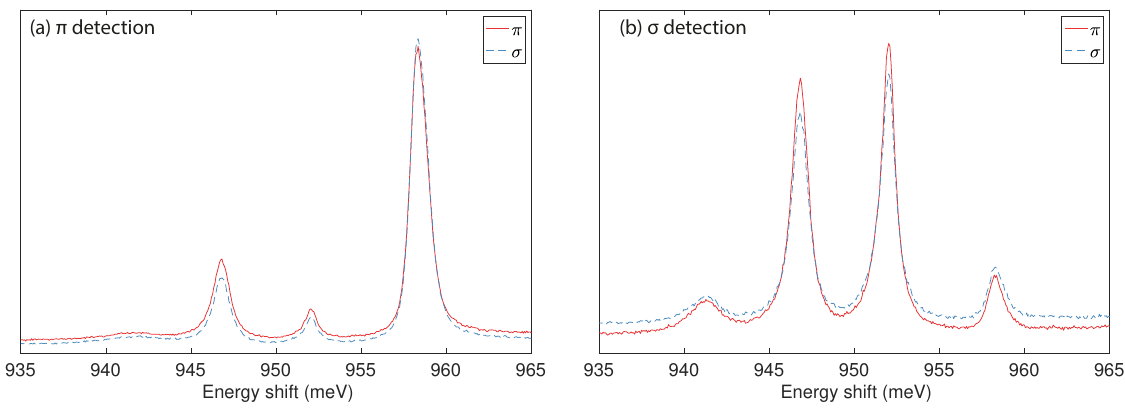}
	\caption{Comparison of raw spectra collected at an applied \hkl<110> stress of \SI{2.1}{\giga\pascal}. Spectra are given for (a) $\pi$ detection polarization and (b) $\sigma$ detection polarization: the two input polarizations are given in each case. No significant difference between input polarizations is visible at this or any other stress value measured.}
	\label{fig:polarization_comparison}
\end{figure}

We have measured \SiVneutral{} in a sample grown by chemical vapour deposition. The sample has faces \hkl<1-10>, \hkl<111> and \hkl<11-2>. Photoluminescence experiments were performed in backscatter geometry i.e. $\mathrm{Z(\psi_e \psi_d)\bar{Z}}$ in Porto notation, where $\psi_e$ and $\psi_d$ are the excitation and detection $E$ vector, respectively [Fig.~\ref{fig:stress_schematic}]. As discussed in the main text, we find no dependence of the spectra on the input polarization $\psi_e$ [Fig.~\ref{fig:polarization_comparison}], and so all spectra are presented for both detection polarizations only. 

Uniaxial stress was applied to the sample using a home-built ram driven by high pressure nitrogen gas and controlled by a Bronkhorst flow controller. The stress cell was mounted into an Oxford Instruments Optistat for low temperature measurements. All measurements were performed using a \SI{785}{\nano\meter} laser (\SI{1.58}{\electronvolt}). The parameters used to generate the model in the main text are given in Table~\ref{tab:parameters}.

\begin{table}[hb]
	\setlength{\tabcolsep}{12pt}
	\centering
	\caption{Model parameter values used to generate the simulation given in the main text. All parameters are in \si{\milli\electronvolt\per\giga\pascal} except $W$, which is given in \si{\milli\electronvolt}.}
	\label{tab:parameters}
	\begin{tabular}[t]{*{9}{>{$}c<{$}}}
		\mathscr{A}_1 & \mathscr{A}_2 & \mathscr{B} & \mathscr{C} & \mathscr{A}_1^{\prime} & \mathscr{A}_2^{\prime} & \mathscr{B} & \mathscr{C} & W\\[0.2em]
		\hline\\[-1em]
		\num{-0.077} & \num{0.93} & \num{-1.0} & \num{-0.24} & \num{0.97} & \num{1.1} & \num{-4.7} & \num{-1.1} & \num{-6.8}\\
	\end{tabular}
\end{table}

\section{Derivation of the stress Hamiltonian solutions}

Let the stress Hamiltonian of A and E states in the absence of coupling be

\begin{equation}
H_{\text{uncoupled}}=\left(
\begin{array}{ccc}
 W+\text{$\alpha^{\prime}$} & 0 & 0 \\
 0 & \alpha +\beta  & \gamma  \\
 0 & \gamma  & \alpha -\beta  \\
\end{array}
\right)\;.
\end{equation}

\noindent The Hamiltonian describing the coupling interaction between the states is

\begin{equation}
H_{\text{coupled}}=\left(
\begin{array}{ccc}
 0 & \gamma^c & \beta^c \\
 \gamma^c & 0 & 0 \\
 \beta^c & 0 & 0 \\
\end{array}
\right)
\end{equation}

\noindent The eigenbasis of the coupling-free $H_{\text{uncoupled}}$ is
\begin{equation}
\left(
\begin{array}{ccc}
 1 & 0 & 0 \\[0.4em]
 0 & \cos\left[\frac{\theta }{2}\right] & -\sin\left[\frac{\theta }{2}\right] \\[0.4em]
 0 & \sin\left[\frac{\theta }{2}\right] & \cos\left[\frac{\theta }{2}\right] \\
\end{array}
\right)
\end{equation}

\noindent Transforming into this basis, the matrix representation of the total Hamiltonian $H=H_{\text{uncoupled}}+H_{\text{coupled}}$ is

\begin{equation}
H=\left(
\begin{array}{ccc}
 W+\alpha^{\prime} & \gamma^c \cos\left[\frac{\theta}{2}\right]+\beta^c \sin\left[\frac{\theta }{2}\right] & \beta^c \cos\left[\frac{\theta }{2}\right]-\gamma^c \sin\left[\frac{\theta }{2}\right] \\[0.4em]
 \gamma^c \cos\left[\frac{\theta }{2}\right]+\beta^c \sin\left[\frac{\theta }{2}\right] & \alpha +\beta  \cos[\theta
]+\gamma  \sin[\theta ] & \gamma  \cos[\theta ]-\beta  \sin[\theta ] \\[0.4em]
 \beta^c \cos\left[\frac{\theta }{2}\right]-\gamma^c \sin\left[\frac{\theta }{2}\right] & \gamma  \cos[\theta
]-\beta  \sin[\theta ] & \alpha -\beta  \cos[\theta ]-\gamma  \sin[\theta ] \\
\end{array}
\right)
\end{equation}

\noindent The expressions for $\alpha$, $\beta$ and $\gamma$ are defined by the symmetry of the center (\DIIId{}), and are given below following \cite{SHughes1967,SDavies1979}:
\begin{align}
  \alpha &= \mathscr{A}_1(\stress_{XX} + \stress_{YY} + \stress_{ZZ}) + 2\mathscr{A}_2(\stress_{YZ}+\stress_{ZX}+\stress_{XY})\nonumber\\
  \beta &= \mathscr{B}(2\stress_{ZZ} - \stress_{XX} - \stress_{YY}) + \mathscr{C}(2\stress_{XY} - \stress_{YZ} - \stress_{ZX}) \label{eqn:model_parameters}\\
  \gamma &= \sqrt{3}\mathscr{B}(\stress_{XX} - \stress_{YY}) + \sqrt{3}\mathscr{C}(\stress_{YZ} - \stress_{ZX})\nonumber
\end{align}

\noindent Here, the $\stress_{ij}$ refer to elements of the stress matrix expressed in the crystal axes. $\alpha^{\prime}$ is defined as $\alpha$ but with $\mathscr{A}_1^{\prime}$, $\mathscr{A}_2^{\prime}$ to reflect the different piezospectroscopic response of the doublet and singlet states. Similarly, $\beta^{c}$ and $\gamma^{c}$ are as $\beta$, $\gamma$ with $\mathscr{B}^{c}$ and $\mathscr{C}^{c}$. $W$ is the difference in energy between the doublet and singlet excited states. The reduced matrix elements $\mathscr{A}_1^{(\prime)}$, $\mathscr{A}_2^{(\prime)}$, $\mathscr{B}^{(c)}$, and $\mathscr{C}^{(c)}$ have the same form as given by \cite{SDavies1976a}.

We now construct the Hamiltonian for each sub-ensemble for each stress direction.

\subsection*{{\texorpdfstring{\hkl<111> stress}{111 stress}}}
The angle between the defect symmetry axis $z$ and the applied stress axis $\uvec{\stress}$ is denoted $\theta_\stress$. For \hkl<111> stress applied to a trigonal defect, we need only consider two cases: the `unique' orientation with $\theta_\stress=\SI{0}{\degree}$; and the three equivalent orientations with $\theta_\stress=\SI{109}{\degree}$. 

The stress matrix is constructed as $\stress_{ij} = \stress(\uvec{\stress}.i)\times(\uvec{\stress}.j)$, where $i, j$ run over the crystal axes $X,Y,Z$, and is subsequently rotated into each orientation frame. For the representative orientations 1 \& 2 [see Table~\ref{tab:orientation_reference}] with the substitution $\theta=\lim_{x\to\beta}\frac{\gamma}{\beta}$, the Hamiltonian parameters are:

\noindent\[\arraycolsep=10pt
\begin{array}{l|c|c|c|c|c|c}
  & \alpha & \beta\equiv\Delta  & \gamma  & \alpha^{\prime} 	& \beta^c\equiv\Omega & 	\gamma^c\\
\hline
 \text{\SI{0}{\degree} sub-ensemble} & \stress(\mathscr{A}_1 + 2\mathscr{A}_2) & 	0 & 0 & \stress(\mathscr{A}_1^{\prime} + 2\mathscr{A}_2^{\prime}) & 0 & 0\\[0.4em]
 \text{\SI{109}{\degree} sub-ensemble} & \stress(\mathscr{A}_1 - \frac{2}{3}\mathscr{A}_2) & 	\frac{4}{3}\mathscr{C}\stress & 0 & \stress(\mathscr{A}_1^{\prime} - \frac{2}{3}\mathscr{A}_2^{\prime}) & \frac{4}{3} \mathscr{C}^{c}\stress & 0\\
\end{array}\]

\noindent Finally, the eigenvalues of the resulting Hamiltonian are as above with $\Delta\equiv\beta$ and $\Omega\equiv\beta^c$:

\noindent\[\arraycolsep=10pt
\begin{array}{l|c|c|c|c}
  & \alpha & \Delta  & \alpha^{\prime}  & \Omega  \\
\hline
\text{\SI{0}{\degree} sub-ensemble} & \stress  \left(\mathscr{A}_1+2 \mathscr{A}_2\right) & 0 & \stress  \left(\mathscr{A}_1^{\prime}+2 \mathscr{A}_2^{\prime}\right)  & 0 \\[0.4em]
\text{\SI{109}{\degree} sub-ensemble} & \stress  \left(\mathscr{A}_1^{\prime}-\frac{2}{3} \mathscr{A}_2^{\prime}\right) & \frac{4}{3} \mathscr{C} \stress & \stress  \left(\mathscr{A}_1^{\prime}-\frac{2}{3}\mathscr{A}_2^{\prime}\right) & \frac{4}{3}\mathscr{C}^c\stress \\
\end{array}\]

\subsection*{{\texorpdfstring{\hkl<110> stress}{110 stress}}}
For \hkl<110> applied stress, we need again only consider two cases: the pair of orientations with $\theta_\stress=\SI{35}{\degree}$; and the pair of orientations with $\theta_\stress=\SI{90}{\degree}$. For the representative orientations 1 \& 3 [see Table~\ref{tab:orientation_reference}], the Hamiltonian parameters are:

\noindent\[\arraycolsep=10pt
\begin{array}{l|c|c|c|c|c|c}
  & \alpha & \beta \equiv \Delta & \gamma  & \alpha^{\prime} 	& \beta^c \equiv \Omega & 	\gamma^c\\
\hline
 \text{\SI{35}{\degree} sub-ensemble} & \stress(\mathscr{A}_1 + \mathscr{A}_2) & \stress(-\mathscr{B}+\mathscr{C}) & 0 & \stress(\mathscr{A}_1^{\prime} + \mathscr{A}_2^{\prime}) & \stress(-\mathscr{B}^c+\mathscr{C}^c) & 0\\[0.4em]
 \text{\SI{90}{\degree} sub-ensemble} & \stress(\mathscr{A}_1 - \mathscr{A}_2) & \stress(-\mathscr{B}-\mathscr{C}) & 0 & \stress(\mathscr{A}_1^{\prime} - \mathscr{A}_2^{\prime}) & \stress(-\mathscr{B}^c-\mathscr{C}^c) & 0\\
\end{array}\]

\noindent As found in the \hkl<111> case, $\Delta\equiv\beta$ and $\Omega\equiv\beta^c$.

\begin{table}[tb]
	\centering
    \caption{The four possible orientations of a trigonal center in a \Tetrahedral lattice.}
    \label{tab:orientation_reference}
    \begin{tabular}[t]{l *{3}{>{$}c<{$}}}
      &x	&	y	&	z\\[0.2em]
      \hline\\[-1em]
      1&\hkl[1-10]	&	\hkl[11-2]	&	\hkl[111]\\
      2&\hkl[-110]	&	\hkl[-1-1-2]	&	\hkl[-1-11]\\
      3&\hkl[110]	&	\hkl[1-12]	&	\hkl[1-1-1]\\
      4&\hkl[-1-10]	&	\hkl[-112]	&	\hkl[-11-1]\\
    \end{tabular}
\end{table}

\section{Intensities of stress-split transitions}

As discussed above and in the main text, for photoluminescence stress measurements performed with an ionizing input beam, the spectra are essentially invariant to input polarization and therefore the expected intensities therefore reduce to the case encountered in absorption measurements.

The expressions for the intensities given in the main text require the intensities of each transition at zero stress in the experimental geometry. The analytical values have been calculated in several places \cite{SDavies1976a,SMohammed1982}. However, the sample used in our experiment has \hkl{111}, \hkl{11-2} and \hkl{1-10} faces: the standard tables give intensities for \hkl<110> or \hkl<001> readout under \hkl<1-10> stress. In Table~\ref{tab:intensities} we give the zero-stress intensities for both \hkl<111> and \hkl<110> stress, including intensities of transitions when measured with detection polarization $\psi_d\|\hkl<11-2>$ under $\vect{\stress}\|\hkl<110>$, as found in our experiment.

\begin{table}[hb]
	\setlength{\tabcolsep}{12pt}
	
	\centering
	\caption{Analytical intensities for different detection polarizations for an $\spinstate{}{E}{}\leftrightarrow\spinstate{}{A}{2}$ transition at a trigonal center, following \cite{SRogers2015}. For \hkl<110> stress, the $\sigma$ polarization values are calculated for a perpendicular direction of \hkl<11-2>, as employed in our experiment.}
	\label{tab:intensities}
	\begin{tabular}{l l @{\hspace{0.6em}} *{5}{l}}
		\text{Stress} 	& \multicolumn{2}{c}{Orientation} & \text{Sym.} & \text{Energy} & $\pi$ & $\sigma$\\
		\multirow{4}{*}{$\hkl<111>$} & \circled{1} & $\SI{0}{\degree}$  & $E_X, E_Y$ & $\mathscr{A}_{1} + 2 \mathscr{A}_2$ & 0 & 1\\[0.8em]
		& \circled{2} && \multirow{1.5}{*}{$E_X\;(\Gamma_1)$} & \multirow{1.5}{*}{$\mathscr{A}_1 - \frac{2}{3}\mathscr{A}_2 + \frac{4}{3}\mathscr{C}$} & \multirow{1.5}{*}{0} & \multirow{1.5}{*}{$\frac{3}{2}$}\\
		& \circled{3} & $\SI{70}{\degree}\;(XZ)$\\
		& \circled{4} && \multirow{-1.5}{*}{$E_Y\;(\Gamma_2)$} & \multirow{-1.5}{*}{$\mathscr{A}_1 - \frac{2}{3}\mathscr{A}_2 - \frac{4}{3}\mathscr{C}$} & \multirow{-1.5}{*}{$\frac{8}{3}$} & \multirow{-1.5}{*}{$\frac{1}{6}$}\\[3em]
		\multirow{4}{*}{$\hkl<110>$} & \circled{1} & \multirow{2}{*}{$\SI{35}{\degree}\;(XZ)$} & $E_X\;(\Gamma_1)$ & $\mathscr{A}_1 + \mathscr{A}_2 - \mathscr{B} + \mathscr{C}$ & 0 & $\frac{2}{3}$\\
		& \circled{2} && $E_Y\;(\Gamma_2)$ &  $\mathscr{A}_1 + \mathscr{A}_2 + \mathscr{B} - \mathscr{C}$ & $\frac{2}{3}$ & $\frac{8}{9}$\\[0.8em]
		& \circled{3} & \multirow{2}{*}{$\SI{90}{\degree}\;(YZ)$} & $E_X\;(\Gamma_1)$ &  $\mathscr{A}_1 - \mathscr{A}_2 - \mathscr{B} - \mathscr{C}$ & 2 & 0\\
		& \circled{4} & & $E_Y\;(\Gamma_2)$ &  $\mathscr{A}_1 - \mathscr{A}_2 + \mathscr{B} + \mathscr{C}$ & 0 & $\frac{10}{9}$
	\end{tabular}
\end{table}

\section{976 nm transition}

\begin{figure}[hbt]
	\centering
	\includegraphics[width=0.6\textwidth]{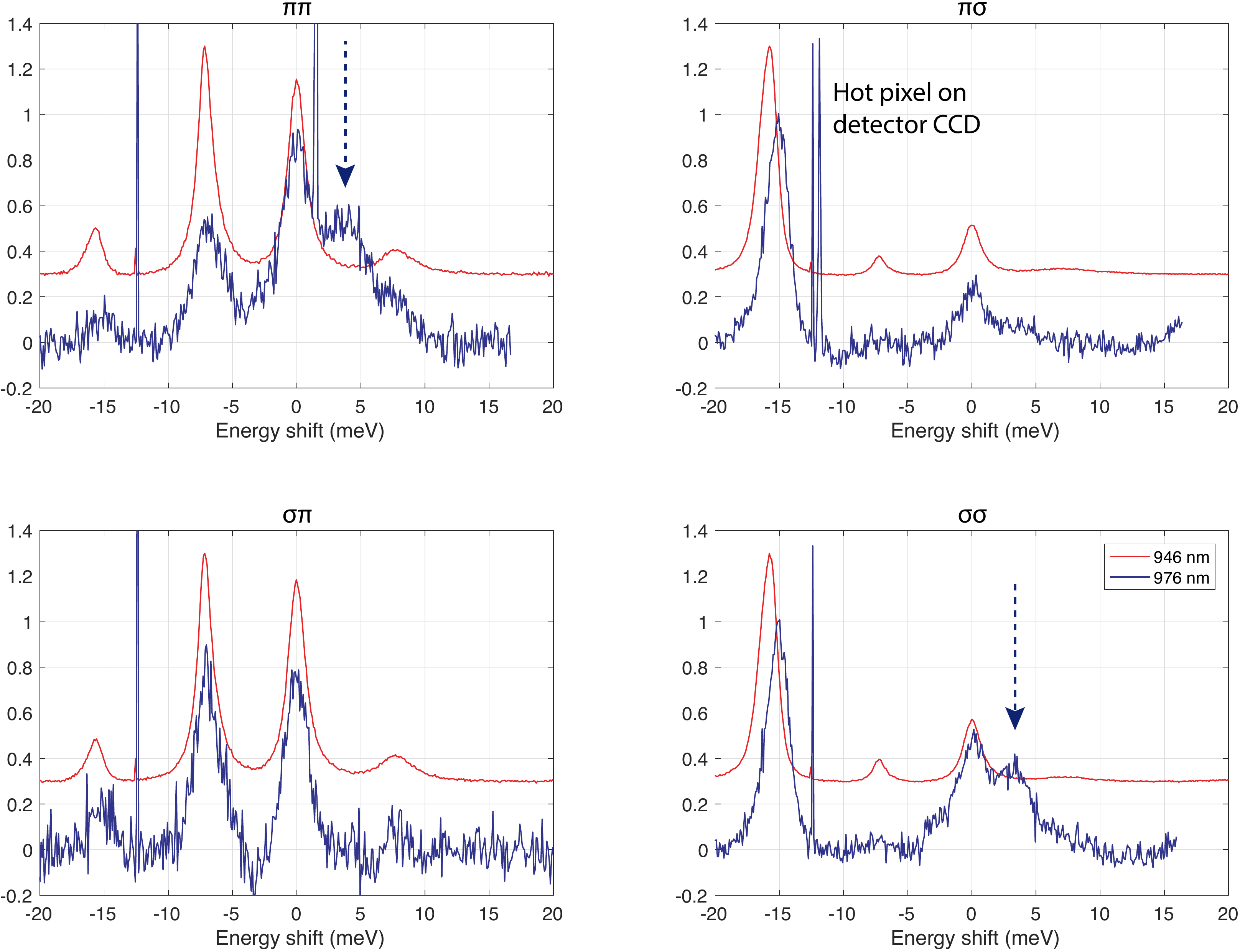}
	\caption{Comparison of \SI{946}{\nano\meter} spectra (red) with \SI{976}{\nano\meter} spectra under \SI{2.1}{\giga\pascal} of applied \hkl<110> stress. The spectra are labelled with excitation and detection polarization. In each case, spectra are essentially identical except for the feature marked with an arrow in the $\pi\pi$ and $\sigma\sigma$ spectra. No other feature of the \SiVneutral{} system is sensitive to input polarization and therefore we assign it to an unrelated defect emitting close to the \SI{976}{\nano\meter} transition.}
	\label{fig:976_hump}
\end{figure}

As described in the main text, the qualitative behavior of the \SI{946}{\nano\meter} and \SI{976}{\nano\meter} transitions is identical. However, a small additional transition appears in certain excitation-detection combinations, namely $\pi\pi$ and $\sigma\sigma$ [Fig~\ref{fig:976_hump}]. As no other features of the \SI{946}{\nano\meter} system are sensitive to input polarization in these measurements, we attribute this additional peak to an unrelated feature.

\section{Spin polarization mechanism}
\begin{figure}[hbt]
	\centering
	\includegraphics[width=0.6\textwidth]{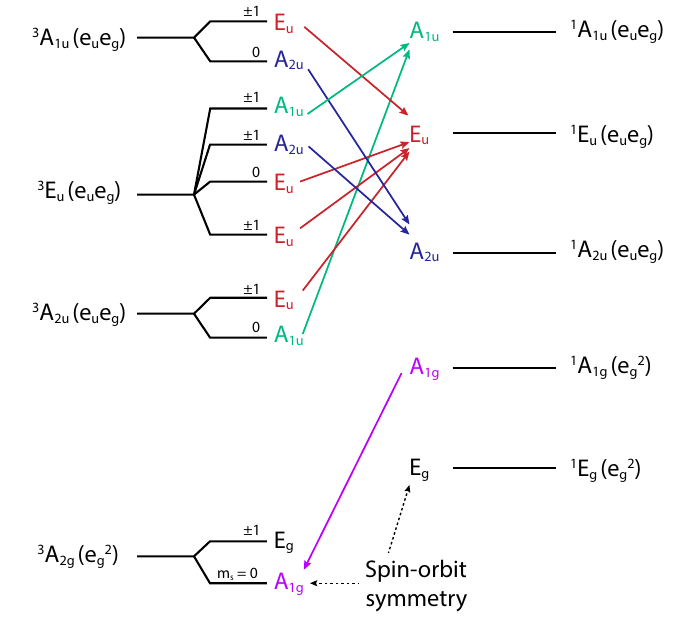}
	\caption{First-order intersystem crossings involving only \vibstate{A}{1g} phonons. The electronic symmetries are given on the far left and right of the figure, with the spin-orbit symmetry given in the center. The states are ordered according to their Coulomb repulsion energy.}
	\label{fig:spin-orbit}
\end{figure}

The electronic structure of \SiVneutral{} is complex, with three and six electronic states arising from the first two lowest-energy electronic configurations \oneelec{e_g^2} and \oneelec{e_ue_g}, respectively. Considering only symmetric \vibstate{A}{1g} phonons, the first-order intersystem crossings (ISC) from the triplet manifold to the singlet manifold are given in Fig.~\ref{fig:spin-orbit}. In this picture, there are no ISCs from the \spinstate{3}{A}{2u} to lower-energy singlet states, suggesting spin polarization should decrease at low temperature, contrary to experiment. Additional information on the relative energy and ordering of the singlets is required for further analysis.

\clearpage

\end{document}